\documentclass[epj]{svjour}
\usepackage{graphicx}
\usepackage{epstopdf}
\usepackage{dcolumn}
\usepackage{bm}
\usepackage{epsfig}
\usepackage{amsfonts}
\usepackage{amssymb,amscd}
\usepackage{hyperref}
\usepackage{xcolor}
\hypersetup{
    colorlinks=true,                
    breaklinks=true,                
    urlcolor= black,                
    linkcolor= blue,                
    bookmarksopen=false,
    filecolor=black,
    citecolor=blue,
    linkbordercolor=blue,
    pdfborder = {0 1 0}
}

\begin{document}

\title{The color dipole picture for prompt photon production in $pp$ and $pPb$ collisions at the CERN-LHC}


\author{
G. Sampaio dos Santos\inst{1} \and
G. Gil da Silveira\inst{1,2} \and
M. V. T. Machado\inst{1}
}

\institute{High Energy Physics Phenomenology Group, GFPAE  IF-UFRGS \\
Caixa Postal 15051, CEP 91501-970, Porto Alegre, RS, Brazil \and
Departamento de F\'{\i}sica Nuclear e de Altas Energias, Universidade do Estado do Rio de Janeiro\\
CEP 20550-013, Rio de Janeiro, RJ, Brazil}

\date{Received: date / Revised version: date}

\abstract{
A study on the prompt photon production within the QCD color dipole picture with emphasis in $pp$ and $pA$ collisions at the LHC energy regimes is performed. We present predictions for the differential cross section as a function of photon transverse momentum at different rapidity bins considering updated phenomenological color dipole models, which take into account the QCD gluon saturation physics. The results are directly compared to the recent experimental measurements provided by CMS and ATLAS Collaborations, showing a reasonable agreement in all rapidity bins with no free parameters. Special attention is given to the IPSAT model given its good description of the data in all rapidity bins from low- to high-$p_{T}$ ranges. As a result, a free-parameter approach has succeeded in describing the LHC data for prompt photon production, while new predictions for the 13-TeV data is presented in view of new data to confirm such prospect.}


\authorrunning{Sampaio dos Santos, Gil da Silveira, and Machado}
\titlerunning{The color dipole picture for prompt photon production at LHC}
\maketitle

\section{Introduction}
\label{intro}

The production of photons in hadronic collisions can be understood as a superposition of different sources of production, and isolation criteria are used to reduce the contamination by photons originating from certain production mechanisms. A photon produced in a hadronic collision is considered prompt when it does not originate from the decay of a hadron, such as $\pi^{0}$ or $\eta$, or when produced with a large transverse momentum, $p_T$. Moreover, the terminology isolated photons concerns to the imposition of  an isolation criterion where a photon is said to be isolated if, in a cone of radius $R$ in rapidity $y^{\gamma}$ and azimuthal angle $\phi^{\gamma}$ around the photon direction, the amount of deposited hadronic transverse energy is smaller than some cut, $E_{\mathrm{cut}}^{\mathrm{had}}=(E_T)^h_{\mathrm{max}}$,  defined by the experiment (i.e., $E_T^h\leq E_{\mathrm{cut}}^{\mathrm{had}}$ inside the region $(y-y^{\gamma})^2+(\phi-\phi^{\gamma})^2\leq R^2$).  Several data sets on prompt photon production have been collected over the years, covering a large domain of center-of-mass energy and also a wide range of photon rapidity and transverse momentum spectrum. For instance, inclusive measurements of prompt photons have been made at hadron colliders by ATLAS \cite{atlas}, CMS \cite{cms}, CDF \cite{cdf}, and D$\emptyset$ \cite{d0} Collaborations, making the comparison between predictions and experimental data a quite meaningful scenario.

In addition, a detailed understanding of prompt photon production is crucial to improve the knowledge both in experimental and theoretical sides. As such, the quantum chromodymamics (QCD) predictions for direct photons constitute an important background in the measurements of diphoton decay channel \cite{bayatian,aad}.
The study of prompt photons is a subject of investigation for a long time and can be related to the deep inelastic scattering (DIS), the Drell-Yan pair production, and the jet production as an important probe of QCD regimes. Due the nature of the quark-photon vertex, measurements of their production cross sections have been proposed as a clean source of information about the QCD dynamics \cite{pasechnik,acharya,gordon,frixione}. Since photons are colorless probes of the dynamics of quarks and gluons and interact electromagnetically only, they escape unchanged through the colored medium created in a high-energy collision. This becomes possible given that they are not sensitive to the QCD induced final-state interactions and hence leave the system without loss of energy and momentum. Therefore, they are considered a powerful probe to investigate the cold nuclear matter effects in the initial stage of the heavy-ion collisions \cite{helenius}. Besides, studies about photon production in quark-gluon plasma (QGP), known as thermal photons, are also available in the literature (e.g., see Refs.~\cite{steffen,rapp}).

From the theoretical side, a treatment in the context of the QCD CD approach \cite{kop} can describe --  within the same framework -- both direct photon and Drell-Yan pair production processes. 
The prompt photon production reaction can be seen in the target rest system, where the production mechanism resembles a bremsstrahlung \cite{kop1}. Therefore, we can apply the CD formalism to describe the radiation processes \cite{kop2}. Such a formulation includes all perturbative and non-perturbative radiation as well as higher-twist contributions. In the CD picture, the phenomenology is based on the universal dipole cross section, fitted to DIS data and successfully describing the DESY-HERA $ep$ data for inclusive and exclusive processes. In high-energy collisions, or very low-$x$ Bjorken variable, nonlinear QCD effects, such as gluon saturation, becomes relevant and should be taken into consideration. The growth of the gluon density at low-$x$ regime can be controlled by gluon recombination effects with a transition region delimited by a $x$-dependent saturation scale, $Q_s(x)$. It is expected that the low-$p_T$ region be able to provide access to the saturation regime and allows to study spin-dependent and spin-averaged gluon densities (PDFs) of hadrons in a kinematic regime where the theoretical uncertainties from usual perturbative QCD (pQCD) are huge. 

Summarizing the recent results on direct photons within the light-cone dipole picture, their azimuthal anisotropy has been identified with an orientation-dependent dipole cross section and it should contribute to the azimuthal asymmetry of direct photons in $pA$ and $AA$ collisions \cite{Kopeliovich:2007fv}. The orientation was given by an off-diagonal unintegrated gluon density (UGD) at leading order (LO) and in Ref.~\cite{Kopeliovich:2007fv} has been modeled through an eikonal-inspired UGD. Recently, a next-to-leading order (NLO) calculation has been performed \cite{Benic:2018hvb,Benic:2019fbj} in the scope of Color Glass Condensate (CGC) formalism and it was found that the contribution of the NLO channel is significantly larger than the LO one at central rapidities at the LHC energies using an UGD for protons based on CGC effective field theory. The similar case for $pA$ collisions in the very same framework has been addressed in Ref.~\cite{Benic:2016uku} (similar analysis also done in Ref.~\cite{Ducloue:2017kkq}). The role played by gluon saturation effects and the value of the anomalous dimension has been analyzed in \cite{Kopeliovich:2009yw} and authors further shown that Cronin enhancement of direct photons can survive at the LHC energy whether nuclear saturation scale acquires large values \cite{Rezaeian:2009it}. The size of finite coherence length (relevant for low energies as at RHIC) has been investigated in Ref.~\cite{Krelina:2016hkr} using the Green function technique which incorporates the color transparency and quantum coherence effects. The seminal work of Ref.~\cite{JalilianMarian:2012bd} treats the azimuthal correlations in photon-hadron production in $pA$ collisions showing the large suppression of the away-side peak in photon-hadron correlations at forward rapidities. Nuclear modification factor, $R_{pA}$, and photon-hadron azimuthal correlations are predicted. That work has promoted a series of further investigations using state-of-art phenomenology concerned to dipole-nucleus interaction (see, e.g., Refs.~\cite{Rezaeian:2016szi,Kovner:2014qea,Kovner:2015rna,Benic:2017znu,vic}). Additional studies on direct photons that take into account other approaches can be found in Refs.~\cite{goharipour,campbell,roy}.	  
In this work, we perform calculations for direct photon production at large and intermediate $p_T$ in a wide rapidity range considering $pp$ and $pA$ collisions at the LHC. We update  previous studies presented in Ref.~\cite{mm}, where semi-analytical expressions for invariant cross section is given for $pp$ and $pA$ collisions. In this context, the role played by the anomalous dimensions, $\gamma_s$, in the transition between the saturation regime and large-$p_T$ (DGLAP-like regime) is clearly identified. In particular, the anomalous dimension at then saturation limit, $\gamma_s\approx 0.76$, is crucial to describe the low and intermediate $p_T$ region whereas the DGLAP limit, $\gamma_s\rightarrow 1$, describes correctly the large-$p_T$ photon spectrum. The situation is similar for the longitudinal structure function \cite{Machado:2005ez} and  multiplicity of charged hadrons \cite{Goncalves:2006yt}. Here, we consider the state-of-art for the  phenomenological models for the dipole-nucleus amplitude including its impact parameter dependence. We investigate the GG approach for nuclear effects as well as the GS property. We believe that this quantitatively measures the theoretical uncertainties present in the invariant cross section in $pA$ collisions. The main quantity of interest in this study is the nuclear saturation scale, $Q_{s,A}$, that defines the onset of unitarity corrections for a nuclear case. There is an uncertainty of the order of 20\% by considering different prescriptions for it and we will use the one extracted from DIS data for $eA$ collision in the context of GS formalism applied to ion targets \cite{armesto}. Such an approach will be directly compared to the calculation using Glauber-Gribov (GG) multiple scattering corrections. 

The paper is organized as follows. In Sec.~\ref{dirph} we start by providing the theoretical information to compute the differential cross section within the QCD CD formalism. Sec.~\ref{res} presents predictions that are compared to the recent measurements focusing in the LHC kinematic regime. In Sec.~\ref{valid} we discuss the validity of the CD approach within the phase-space region probed in the experimental data. Finally, in Sec.~\ref{conc} we summarize the main conclusions and propose future investigations.

\section{Theoretical framework}
\label{dirph}

\begin{figure}[t]
\begin{center}
\includegraphics[scale=.5]{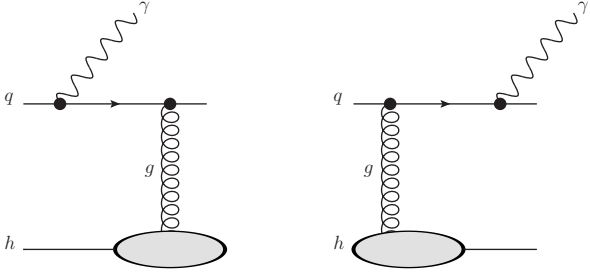}  
\end{center}
\caption{Typical diagrams for real photon bremsstrahlung by a quark (antiquark) interacting with the target via gluon exchange. The photon radiation may happen either before (left panel) or after (right panel) the quark-target scattering.}
\label{bremss}
\end{figure}

In this work we consider the real photon production off protons and nuclear targets at high energies, where the CD system is adopted to describe this mechanism. 
The emission of real photons is then treated as electromagnetic bremsstrahlung by a quark projectile, which interact with the color field of the target in the single gluon approximation, as seen in Fig.~\ref{bremss}, with a photon emitted either before or after the quark-target interaction. At the high energy limit, each of the diagrams in Fig.~\ref{bremss} is factorized into a vertex of the real photon production associated with the quark-target scattering amplitude, which takes part in the matrix element squared \cite{kop1,kop}. Hence, the real photon radiation process can be interpreted in terms of $q\bar{q}$ dipole scattering off the target.

Considering the target as a proton, in Ref.~\cite{kop1} the differential cross section in terms of the photon transverse momentum $p_{T}$ is presented, taking the form
\begin{eqnarray}
\frac{d \sigma(qp\to q\gamma)}{d(\ln \alpha)\,d^{2}\vec{p}_{T}}&=&\frac{1}{(2\pi)^{2}}
\sum_{in,f}\sum_{L,T}
\int d^{2}\vec{r}_{1}d^{2}\vec{r}_{2}e^{i \vec{p}_{T}.(\vec{r}_{1}-\vec{r}_{2})}\nonumber \\
&\times& \phi^{\star T,L}_{\gamma q}(\alpha, \vec{r}_{1})
\phi^{T,L}_{\gamma q}(\alpha, \vec{r}_{2}) \nonumber \\
&\times & \frac{1}{2}\Big[\sigma_{dip}(x,\bar{r}_{1})+\sigma_{dip}(x,\bar{r}_{2}) -\sigma_{dip}(x,\Delta \bar{r})\Big].
\label{dip}
\end{eqnarray}
After evaluated the integration over the final quark kinematics, only two radiation amplitudes contribute to the cross
section, where $\vec{r}_{1}$ and $\vec{r}_{2}$ are the quark-photon transverse
separations entering in $\sigma_{dip}$. Moreover, the transverse displacements of the final quarks in the amplitudes are correspondingly $\bar{r}_1=\alpha r_{1}$ and $\bar{r}_2=\alpha r_{2}$ [with $\Delta \bar{r}=(\bar{r}_1-\bar{r}_2)$]. 
The parameter $\alpha$ is the relative fraction of the quark momentum carried by the photon.
The Bjorken variable $x_{1,2}$ is related to the projectile and target momenta, $x_{1,2} = \frac{p_T}{\sqrt{s}}e^{\pm y^{\gamma}}$, where $y^{\gamma}$ is the photon rapidity and $\sqrt{s}$ is the collision center-of-mass energy. The light-cone wave function of the photon bremsstrahlung is given by
\begin{eqnarray}
&&\sum_{in,f}\phi^{T\star}_{\gamma q}(\alpha, \vec{r}_{1})\phi^{T}_{\gamma q}(\alpha, \vec{r}_{2})
= \frac{\alpha_{em}}{2\pi^{2}} \,\left\{m^2_{q}\alpha^{4}K_{0}(\epsilon r_{1})K_{0}(\epsilon r_{2}) \right. \nonumber \\
&+& \left. [1+(1-\alpha)^{2}]\epsilon^{2}\frac{\vec{r}_{1}.\vec{r}_{2}}{r_{1}r_{2}}
K_{1}(\epsilon r_{1})K_{1}(\epsilon r_{2})\right\},
\label{wave}
\end{eqnarray}
where $K_{0,1}(x)$ are the modified Bessel function of the second kind. The auxiliary variable $\epsilon^{2}=\alpha^{2} m_{q}^{2}$ depends on the effective quark mass, assumed to be $m_{q}=0.2$~GeV in our numerical calculations.

The hadronic cross section is obtained from the convolution of the elementary partonic cross section, Eq.~(\ref{dip}), with the projectile structure function $F_{2}^{p}$ \cite{kop,kop3},
\begin{eqnarray}
\frac{d \sigma (pp\to \gamma X)}{dy^{\gamma}d^{2}\vec{p}_{T}}= \int_{x_{1}}^{1}\frac{d\alpha}{\alpha} F_{2}^{p}\Big(\frac{x_{1}}{\alpha},\mu^2\Big)\, \frac{d \sigma(qp\to q\gamma)}{d(\ln \alpha)\,d^{2}\vec{p}_{T}},
\label{xsec}
\end{eqnarray}
where $\mu^{2} = p^{2}_{T}$ will be considered and a $F_{2}^{p}$ parametrization presented in Ref.~\cite{adeva}. The Fourier integrals over $\vec{r}_{1}$ and $\vec{r}_{2}$ can be simplified to a one-dimensional integral over the dipole separation $r$, which was first derived in Ref.~\cite{Kopeliovich:2001hf},
\begin{eqnarray}
\frac{d\sigma\,(pp\to \gamma
X)}{dy^{\gamma}d^{2}\vec{p}_{T}} & = &
\frac{\alpha_{em}}{2\pi^2}\int_{x_{1}}^{1}\frac{d\alpha}{\alpha}
 F_{2}^{p}\left(\frac{x_{1}}{\alpha},\mu^2\right) \nonumber\\
&\times & \left\{ m_q^2\alpha^4\left[\frac{{\cal I}_1}{(p_T^2+\varepsilon^2)}-\frac{ {\cal I}_2}{4\varepsilon} \right]
  +  [1+(1-\alpha)^2]\right.\nonumber \\
  &\times& \left. \left[ \frac{\varepsilon p_T \, {\cal I}_3}{(p_T^2+\varepsilon^2)} -\frac{{\cal I}_1}{2}+\frac{\varepsilon \,{\cal I}_2}{4}\right]
\right \}.
\label{hank}
\end{eqnarray}
The quantities ${\cal I}_{1,2,3}$ are Hankel integral transforms of order $0$ (${\cal I}_{1,2}$) and order $1$ (${\cal I}_{3}$) given by:
\begin{eqnarray}
\label{hankel1}
{\cal I}_1 & = & \int_0^{\infty}dr\,rJ_0(p_T\,r)K_0(\varepsilon\,r)\,\sigma_{dip}(x_2,\alpha r),\\
{\cal I}_2  &=&  \int_0^{\infty}dr\,r^2J_0(p_T\,r)K_1(\varepsilon\,r)\, \sigma_{dip}(x_2,\alpha r), \\
{\cal I}_3 & = & \int_0^{\infty}dr\,rJ_1(p_T\,r)K_1(\varepsilon\,r)\, \sigma_{dip}(x_2,\alpha r).
\label{hankel3}
\end{eqnarray}
In the color transparency region,  $\sigma_{dip}\propto r^2$, the Hankel integrals can be analytically computed, resulting in:
\begin{eqnarray}
{\cal I}_1  &\propto& \frac{(\varepsilon^2-p_T^2)}{(p_T^2+\varepsilon^2)^3},\\
{\cal I}_2 &\propto& \frac{4\varepsilon\,(\varepsilon^2-2p_T^2)}{(p_T^2+\varepsilon^2)^4},\\
{\cal I}_3 &\propto& \frac{2p_T\varepsilon}{(p_T^2+\varepsilon^2)^3},
\label{aproxis}
\end{eqnarray}
where the exact prefactors for GBW model (with $\gamma_{\mathrm{eff}}=1$) can be found in Ref.~\cite{mm}. Notice that in the absence of saturation, the CD approach can be related to the QCD Compton process as demonstrated in Refs.~\cite{field,Betemps:2003je}.

For our purposes, we consider here some phenomenological models based on the idea of parton saturation in order to investigate the differences and uncertainties among them. In a general form, the dipole-proton cross section can be parametrized as follows
\begin{eqnarray}
\label{sigdip}
\sigma_{dip}(x,\vec{r};\gamma) &=&\sigma_{0}\left[ 1-\exp\left(-\frac{r^{2}Q_{s}^{2}}{4}\right)^{\gamma_{\mathrm{eff}}}\,\right],\\
 Q_{s}^2(x) &=& \left(\frac{x_0}{x}\right)^{\lambda},
\label{param}
\end{eqnarray}
where $\gamma_{\mathrm{eff}}$ stands for the effective anomalous dimension and $Q_{s}$ is the saturation scale. For instance, in Golec-Biernat-Wüsthoff (GBW) saturation model \cite{gbw} one has $\gamma_{\mathrm{eff}} = 1$ and fitting parameters \cite{gbwfit} using four quark flavors assuming the values $\sigma_0 = 27.32$ mb, $x_0 = 0.42\times 10^{-4}$, and $\lambda = 0.248$. Another model that has the same form as Eq.~(\ref{param}) is the Boer-Utermann-Wessels (BUW) model \cite{buw}. In this model the effective anomalous dimension takes the form,
\begin{eqnarray}
\gamma_{\mathrm{eff}}= \gamma_{s}+(1-\gamma_{s})\frac{(\omega^a-1)}{(\omega^a-1)+b},
\label{buw}
\end{eqnarray}
where $\omega\equiv p_T/Q_{s}$ and the free parameters are given by $a=2.82$ and $b=168$ obtained from a fit to describe the RHIC data on hadron production. One common characteristic is that, for large $p_T$, the dipole cross section in the BUW model reproduces the GBW predictions by using a different set of fitting parameters: $\gamma_{s} = 0.63$, $\sigma_0 = 21$ mb, $x_0 = 3.04\times 10^{-4}$, and $\lambda = 0.288$.

In high-energy collisions (or equivalent low-$x$ regime) the effects of QCD parton evolution are present; in particular the effects coming from multiple parton scattering. In order to analyze the effect of QCD evolution in the dipole cross section, we add to our studies the Impact Parameter Saturation (IPSAT) model \cite{ipsat}. In this case, the dipole cross section depends on a gluon distribution evolved via DGLAP equation:
\begin{eqnarray} 
\sigma_{dip} (x,\vec{r}) &=& 2\int d^2b\,N(x,r,b),\\
N(x,r,b)& = & 1-\exp\left(-\frac{\pi^2}{2N_c}r^2\alpha_S(\mu^2)xg(x,\mu^2)T(b)\right),\nonumber
\label{ipsat1}
\end{eqnarray}
where $N$ is the dipole-nucleon scattering amplitude with a factorized impact-parameter dependence given by a Gaussian profile, $T(b)$, for the proton
\begin{eqnarray}
T(b) = \frac{1}{2\pi B_{G}}  
\exp\left(-\frac{b^{2}}{2B_{G}} \right).
\label{profile}
\end{eqnarray}
The initial gluon distribution has the form, $xg(x,\mu_{0}^{2}) =  A_{g}x^{-\lambda_{g}} (1-x)^{6}$, which is evolved from a scale $\mu_0^2$ up to $\mu^2$ using the DGLAP evolution equations without quarks, with $\mu^2=4/r^2+\mu_0^2$ related to the dipole size $r$. The parameters are extracted from a fit to high-precision combined HERA data for the reduced cross section (see Ref.~\cite{ipsatfit}). 

One of the goals of this work is to estimate the cross section for prompt photon production in proton-nucleus collisions, where $A$ is the nucleus atomic mass number. Within the QCD CD picture, there are basically two ways to implement the nuclear effects: GS property from parton saturation models and GG formalism for nuclear shadowing. We refer to Ref.~\cite{armesto} as an example of using GS to include $A$-dependence in the  scattering cross section. There, the authors have studied how experimental data on lepton-nucleon collisions constrain characteristic features of particle production in nuclear collisions, such as their dependence on $\sqrt{s}$ and on $A$. They have demonstrated that the cross section for DIS off nuclei,  $\gamma^*A\rightarrow X$, can be written in terms of the cross section for DIS off nucleons,  $\gamma^*p\rightarrow X$, assuming a dependence only on the scaling variable $\tau=Q^2/Q_s(x)$ instead of $x$ and $Q^2$ separately. Then, the nuclear effects are absorbed into the saturation scale and on nucleus transverse area, $S_A=\pi R_A^2$ (compared to the nucleon one, $S_p=\sigma_0/2=\pi R_p^2$). Here, we assume that GS is valid in the dipole-nucleus amplitude, $N_A$, and, consequently, this is translated into a $A$-dependence on prompt photon production cross section, 
\begin{eqnarray}
\frac{\sigma(pA\to \gamma X)}{S_A} = \frac{\sigma(pp\to \gamma X)}{S_p},
\label{prescr}
\end{eqnarray}
being the saturation scaling in protons, $Q_s$, replaced by a nuclear scaling, $Q_{s,A}$, in the following way:
\begin{eqnarray}
Q_{s,A}^2&=&Q_{s,p}^2\left(\frac{A \pi R_p^2}{\pi R_A^2}\right)^\frac{1}{\delta}, \\
N_A(x,r,b) & = & N(rQ_{s,p}\rightarrow rQ_{s,A}),
\label{qs2A}
\end{eqnarray}
which grows with the quotient $1/\delta$. The expression for the nuclear radius is $R_A=(1.12 A^{1/3}-0.86A^{-1/3})$~fm, while the 
$\delta$ and $\pi R_p^2$ are parameters determined by data, resulting in $\delta=0.79$ and $\pi R_p^2=1.55$~fm$^2$ \cite{armesto}. The very same ansatz has been considered also to describe data for exclusive vector meson production and DVCS at DESY-HERA as well as photonuclear $\gamma A$ cross section in meson production extracted from ultraperipheral collisions at the LHC \cite{Ben:2017xny}. 

On the other hand, we can use GG formalism to write the dipole-nucleus scattering cross section in terms of the nuclear profile:
\begin{eqnarray}
\label{ipsat2} 
\sigma_{dip}^{nuc} (x,\vec{r};A) &=& 2\int d^2b\,N_A(x,r,b),\\\nonumber
N_A(x,r,b)& = & 1-\exp\left(-\frac{\pi^2}{2N_c}r^2\alpha_Sxg(x,\mu^2)T_A(b)\right), \\
\end{eqnarray}
with the thickness function, $T_A$, computed from the Woods-Saxon distribution. 

In next section we will use these phenomenological models to compute the $p_T$ and $y^{\gamma}$ distributions of direct photon production in $pp/pA$ collisions at the LHC.

\section{Numerical results and discussions}
\label{res}

Let us present the predictions obtained with the QCD CD framework for prompt photon production in $pp$ and $pA$ collisions. We estimate the transverse momentum and rapidity distributions focusing at the LHC energies and using three phenomenological models for the dipole cross section discussed in the previous section (GBW, BUW, and IPSAT) with the corresponding introduction of nuclear effects via GS and GG shadowing.

Before comparing the theoretical predictions to the  experimental results some comments are in order. Experimentally, an isolation
cut is applied and, in our case, an isolation cone with radius 
$R=\sqrt{(\eta^q-\eta^{\gamma})^2+(\phi^q-\phi^{\gamma})^2}$ around the photon direction would need to be considered. Here, $\eta^q$ and $\phi^q$ are the pseudorapidity and azimuthal angle of the final state (anti) quark, respectively. Both CMS and ATLAS Collaborations use an isolation radius $R = 0.4$ and maximum hadronic energy $E_h<4-5$~GeV. The present approach takes into account only the direct contribution to the prompt photon production and does not include the fragmentation contribution (it is estimated to be a 10\% contribution at midrapidities for the LHC energies \cite{dEnterria:2012kvo}). Moreover, in Eq.~(\ref{dip}) the integration over final state quark momentum has already been performed (no constraint on the quark rapidity or transverse momentum is imposed). It is expected that the isolation cut would introduce small modifications in the $p_T$-spectra (this is assumed in Refs.~\cite{kop,Kopeliovich:2007fv,Kopeliovich:2009yw,Rezaeian:2009it,Krelina:2016hkr,vic,mm,Klein-Bosing:2014uaa}). The key point is that the isolation cut in usual pQCD modifies the high order (HO), ${\cal{O}}\sim \left(\frac{\alpha_s(\mu^2)}{\pi}\right)^2$, part the direct contribution. The Born (Compton process) term for the direct contribution of order $\left(\frac{\alpha_s(\mu^2)}{\pi}\right)$, remains unchanged by the cut (this is explicitly shown in Eq.~(5.2) and Table 1 of Ref.~\cite{Catani:2002ny}). As a function of the photon $p_T$, the isolation cut has a small effect on the direct contribution, since it does not act at the Born level and the effect of isolation on the total contribution to the NLO cross section (direct+fragmentation) depends only weakly on $p_T$ (it is around a 10\% correction to the direct contribution and 15\% to the total one, see Fig.~3 of Ref.~\cite{Catani:2002ny}). On the other hand, it has been shown in Refs.~\cite{Betemps:2003je,Raufeisen:2002zp} that, when saturation effects are neglected, the CD approach reproduces the very same QCD Compton contribution in which the quark comes from the projectile and the gluon from the target. At the same time, the resummation of contributions of all orders $\sim [\alpha_s\ln(1/x_2)]^n$ is taken into account and a finite $p_T$-spectrum at $p_T \rightarrow 0$ is obtained if saturation is present in the dipole-target amplitude. Finally, the parametrization for $F_{2}^{p}$ based on Ref.~\cite{adeva} present a detailed list of parameters employed in the fit for $F_{2}^{p}$. As seen in Tab.~XII of Ref.~\cite{adeva}, the uncertainties on the fit parameters are rather small, roughly $\lesssim$5\% and does not impose significant uncertainty on our results presented in this work.

\begin{figure*}[t]
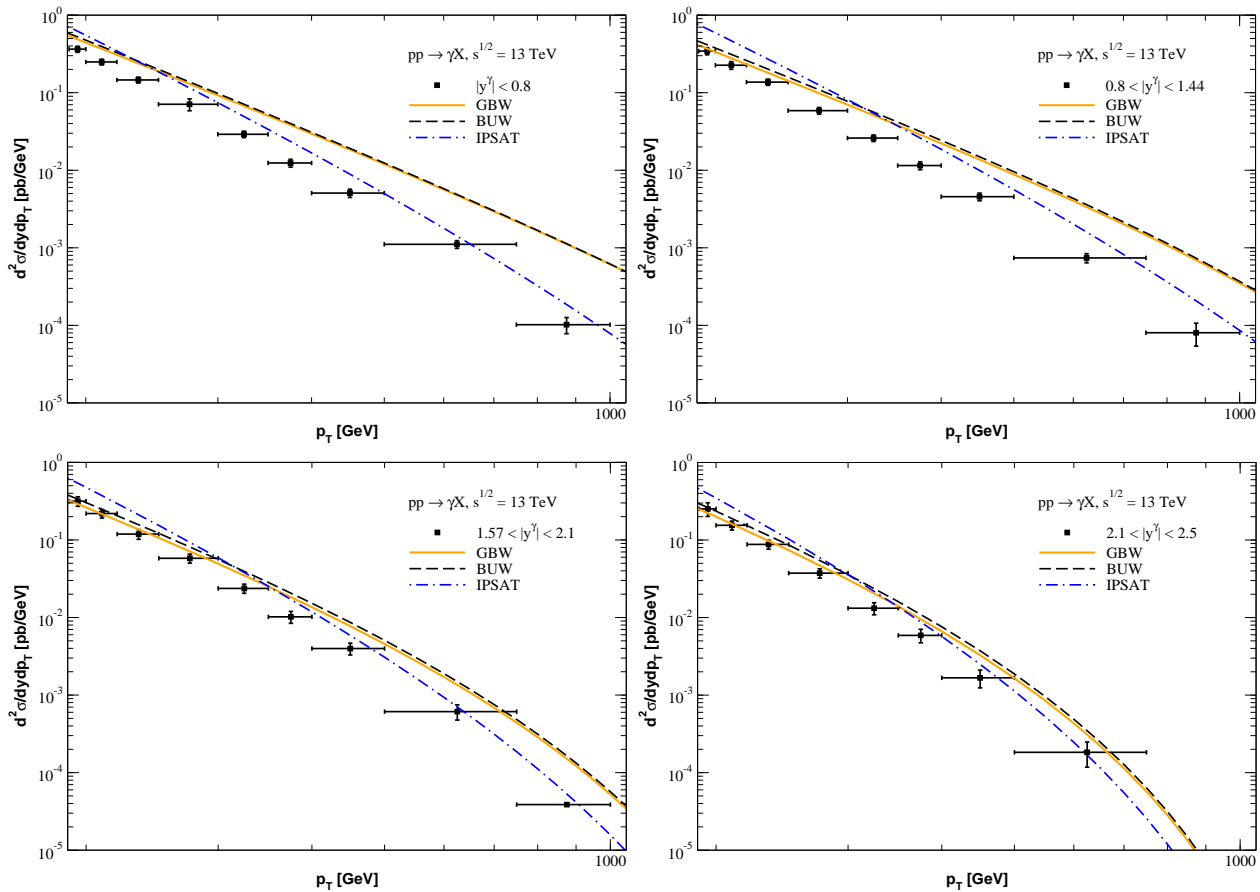

\begin{tabular}{cc}
\includegraphics[scale=0.35]{prompt_pp13_CMS1.eps}  \includegraphics[scale=0.35]{prompt_pp13_CMS2.eps} \\
\includegraphics[scale=0.35]{prompt_pp13_CMS3.eps}  \includegraphics[scale=0.35]{prompt_pp13_CMS4.eps}
\end{tabular}
\caption{Differential cross sections for prompt photon production in $pp$ collision at $\sqrt{s} = 13$~TeV and considering four rapidity bins. The predictions are obtained using three phenomenological CD models and compared to the experimental data from CMS experiment \cite{sirunyan}.}
\label{ppCMS}
\end{figure*}

First, we present the numerical results for $pp$ collisions at $\sqrt{s} = 13$~TeV. Figure~\ref{ppCMS} shows the predictions for the inclusive prompt photon cross sections compared to the measurements from the CMS Collaboration \cite{sirunyan}. The results for the differential cross section as a function of $y^{\gamma}$ and $p_T$ are computed considering four different rapidity bins: $|y^{\gamma}| < 0.8$, $0.8 <|y^{\gamma}| < 1.44$, $1.57 <|y^{\gamma}| < 2.1$, and $2.1 <|y^{\gamma}| < 2.5$. The GBW (solid lines) and BUW (dashed lines) models predict slightly different results in $p_T < 300$~GeV. Apparently, the GBW and BUW models improve the data description in these $p_T$ domain, however we can not distinguish between the models. 
On the other hand, taking $p_T > 300$~GeV, the GBW results are similar to the BUW model as expected, since at large $p_T$ the effective anomalous dimension is identical in both models, namely $\gamma_{\mathrm{eff}} = 1$. The IPSAT results (dot-dashed lines) at $p_T < 300$~GeV are in accordance with the GBW and BUW models, however, as $p_T$ increases, the IPSAT model is in better agreement with data. This improvement compared to the GBW and BUW models comes from the QCD evolution in $\mu^2=p_T^2$ present in the IPSAT model. It does a better job at forward rapidities where smaller values of $x$ are probed. However, at very forward rapidity, the GBW and BUW models are able to predict the correct shape and normalization of the $p_T$-spectrum.  In the calculations using IPSAT, we are using the small-$r$ limit for the dipole-proton amplitude where the Hankel transform can be analytically solved, Eq.~(\ref{aproxis}). This is justified by the fact that the typical dipole sizes being probed in direct photons are $r \propto 1/p_T$, which is sufficiently small at large $p_T$ considered here.

\begin{figure*}[t]
\begin{tabular}{cc}
\includegraphics[scale=0.35]{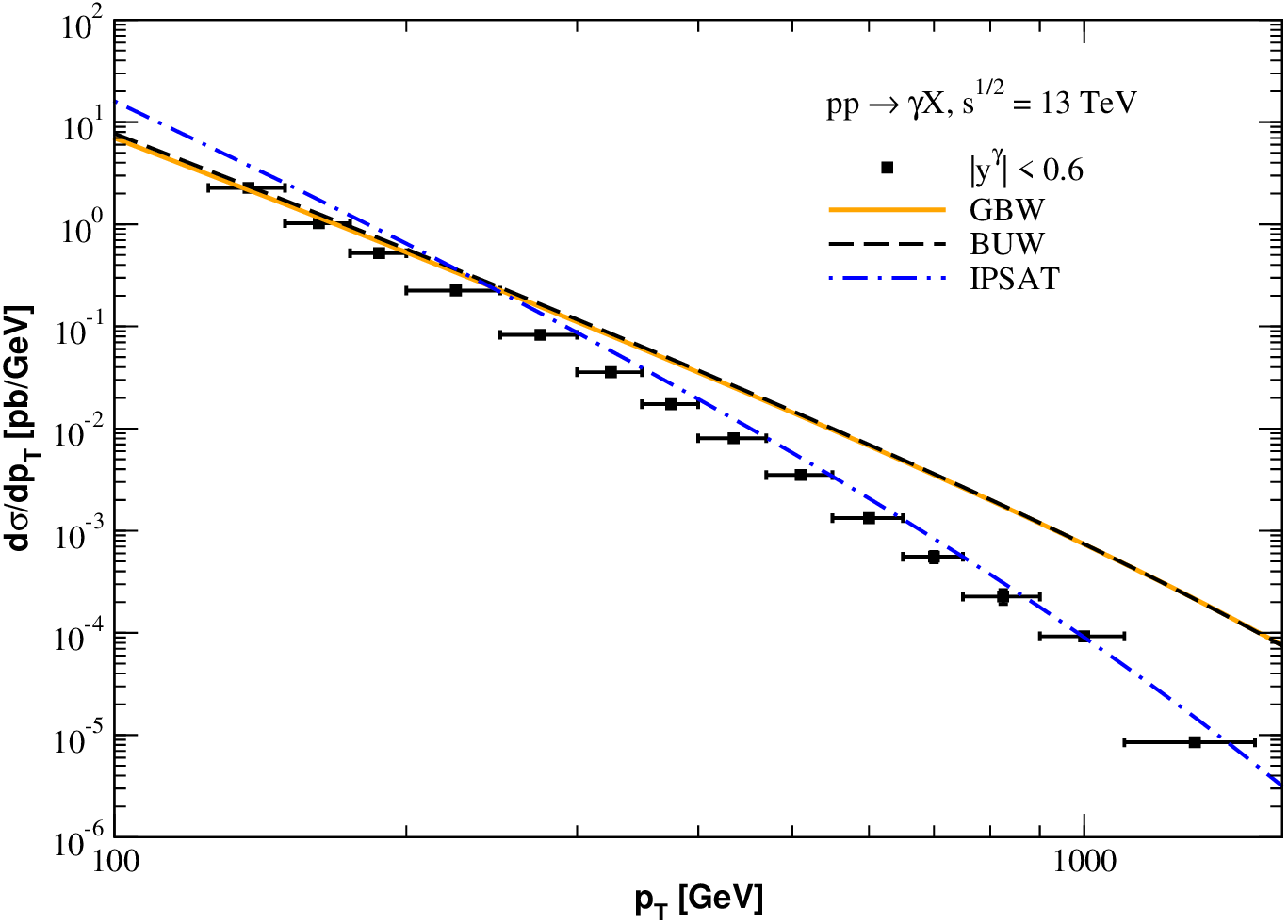}  \includegraphics[scale=0.35]{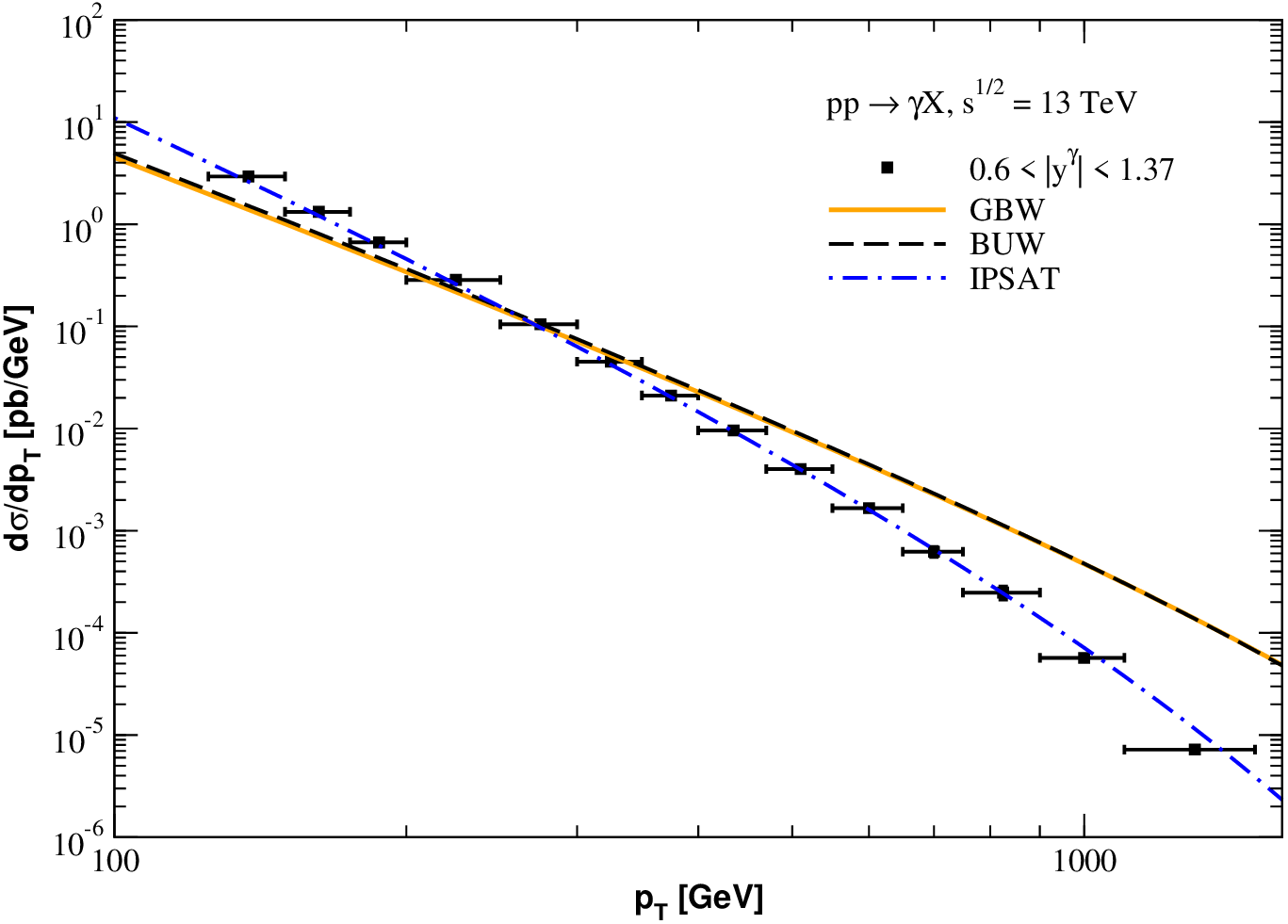} \\
\includegraphics[scale=0.35]{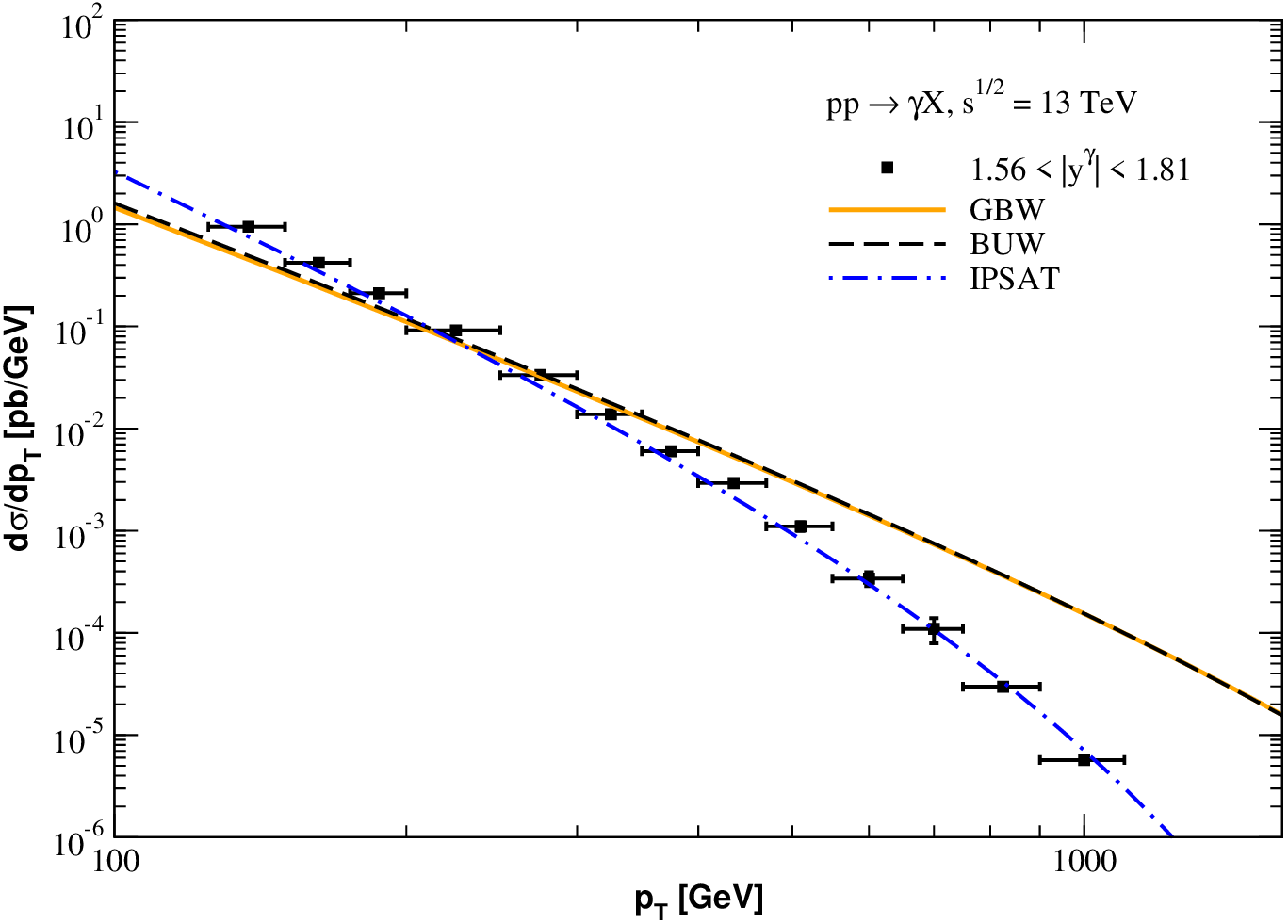}  \includegraphics[scale=0.35]{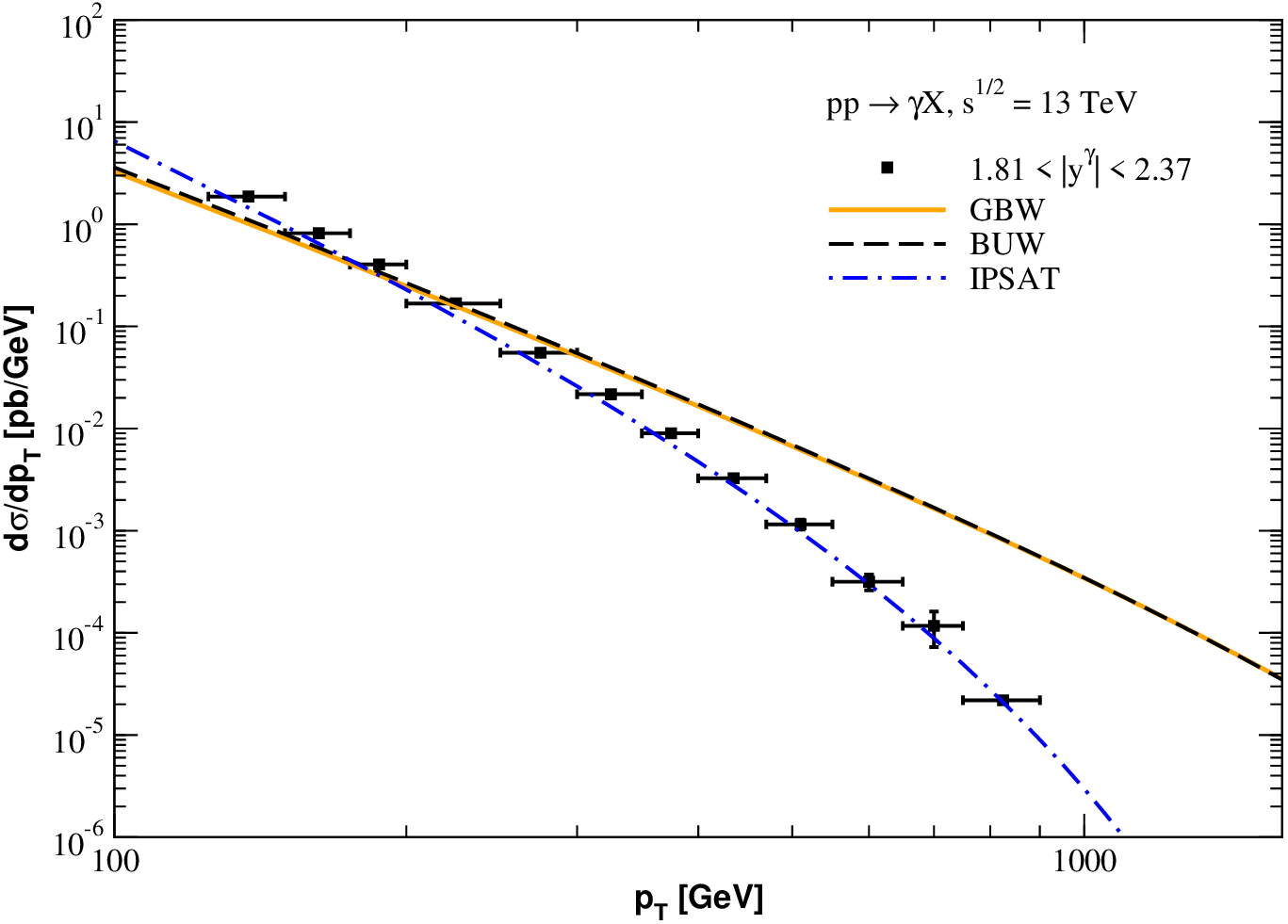}
\end{tabular}
\caption{Differential cross sections for prompt photon production in $pp$ collision at $\sqrt{s} = 13$~TeV and considering four rapidity bins. The predictions are obtained using three phenomenological CD models and compared to the experimental data from ATLAS experiment \cite{aaboud}.}
\label{ppATLAS}
\end{figure*}

In Fig.~\ref{ppATLAS} the predictions are compared to the experimental data from the ATLAS Collaboration \cite{aaboud}. The corresponding results for the differential cross section in terms of $p_T$ are obtained considering four distinct rapidity bins: $|y^{\gamma}| < 0.6$, $0.6 <|y^{\gamma}| < 1.37$, $1.56 <|y^{\gamma}| < 1.81$, and
$1.81 <|y^{\gamma}| < 2.37$. We have verified that we can not distinguish among the results for the three dipole cross section models at the kinematic range of $p_T \lesssim 300$~GeV. Furthermore, the GBW and BUW models overshoots the experimental data beyond $p_T > 300$~GeV. As seen in the CMS data, the IPSAT model provides a good description of the ATLAS data in all rapidity bins, especially a better agreement at large $p_T$. The general conclusion is that color dipole models are able to describe the LHC data at forward rapidities, even at the large $p_T$ range.

Based on the points raised in the present discussion, we propose a simple parametrization for the invariant cross section assuming color transparency in the dipole-target cross section and a DGLAP-like anomalous dimension, $\gamma_{\mathrm{eff}}=1$. This allows to compute analytically the Hankel transforms in Eqs.~(\ref{hankel1}-\ref{hankel3}) and in the massless quark limit, $m_1\rightarrow 0$, only the second term in Eq.~(\ref{hank}) survives. Specifically, the non-vanishing contribution in the second term is proportional to an analytical function: ${\cal I}_1\propto \sigma_0 (\alpha Q_s)^2/p_T^4$. Also, we can write a rough approximation for the nucleon structure function based on the GBW model,
\begin{eqnarray}
F_2(x,Q^2)\approx\frac{\sigma_0 Q^2}{4\pi^2\alpha_{em}}\left(\frac{Q_s^2(x)}{Q^2}\right)^{\gamma_{\mathrm{eff}}}.
\end{eqnarray}
Therefore, the integration over $\alpha$ can be done, obtaining
 \begin{eqnarray}
 \label{dsdpapprox}
\frac{d\sigma\,(pp\to \gamma
X)}{dy^{\gamma}d^{2}\vec{p}_{T}} & \approx & \bar{\sigma}\left(\frac{Q_s^2(x_1)}{p_T^2}\right)\left(\frac{Q_s^2(x_2)}{p_T^2}\right)f(x_1),\\
f(x_1) & \approx & \frac{1012}{1989} -\frac{4}{17}x_1^{\frac{17}{4}}+\frac{8}{13}x_1^{\frac{13}{4}}-\frac{8}{9}x_1^{\frac{9}{4}},
\end{eqnarray}
where $\bar{\sigma}\sim \sigma_0^2/(64\pi^4)\simeq 0.31$~mb/GeV$^2$ and $f(x_1)$ is a well behaved function of $x_1$ resulting from $\alpha$-integration, which is basically a constant for small $x_1$, $f(x_1\ll1)=0.509$ (using $\lambda=0.248\approx 1/4$). That limit occurs, for instance, at central rapidities, $x_1=x_2=p_T/\sqrt{s}$.  This scaling function closely resembles the universal multiplicity scaling for prompt photons investigated in Refs.~\cite{Klein-Bosing:2014uaa,Praszalowicz:2018vfw,Khachatryan:2019uqn}, in which photon $p_T$-spectra at low transverse momentum are scaled with charged hadron pseudorapidity density at midrapidity. In terms of $p_T$ and rapidity $y^{\gamma}$, Eq.~(\ref{dsdpapprox}) results $d\sigma /dy^{\gamma}dp_T \propto (\sqrt{s}/p_T)^{2\lambda}\,f(y,p_T)/p_T^3$. 

\begin{figure*}[t]
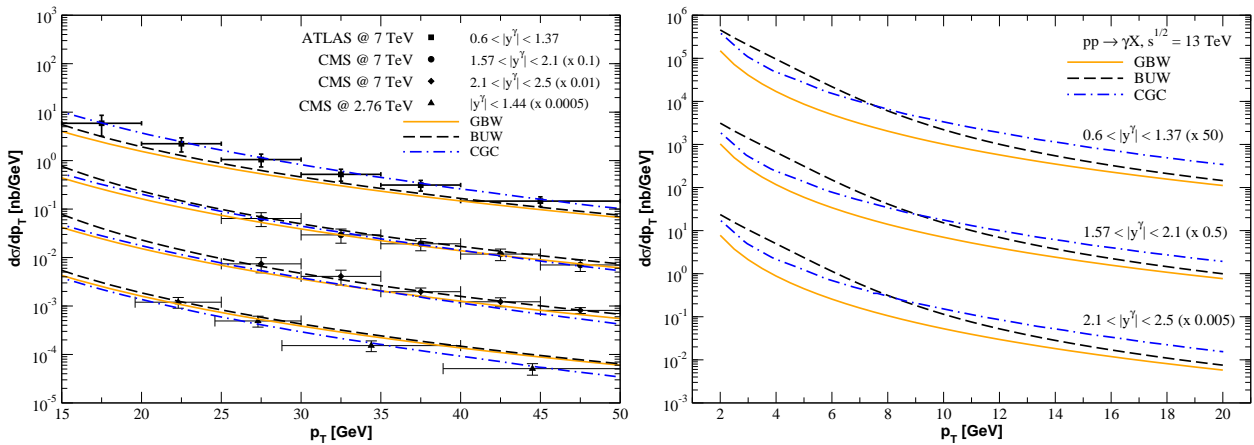

\begin{tabular}{cc}
\includegraphics[scale=0.35]{prompt_pp.eps}
\includegraphics[scale=0.35]{prompt_pp13.eps}
\end{tabular}
\caption{Predictions of the CD models in four different rapidity ranges at a low-$p_{T}$ region (left) compared to the data collected at 2.76 TeV (CMS) and 7~TeV (CMS and ATLAS). In view of future data-takings,  the predictions at 13~TeV are shown for a lower $p_{T}$ range (right) in three rapidity bins.}
\label{fig:comparpp}
\end{figure*}

To evaluate our predictions with the CD model, it is important to compare our results to recent calculations in the literature for the low-$p_T$ region. One of them is the full NLO computation of direct photon cross section in the CGC effective field theory presented in Ref.~\cite{Benic:2018hvb,Benic:2019fbj} (using UGD obtained from CGC formalism) considering energies of 2.76, 7, and 13~TeV. There, authors estimate a 15\% systematic uncertainty in the calculations across several rapidity bins and an overall normalization factor $K=2.4$ was used. Here, the kinematic phase-space in the region where the saturation corrections should be very important behaves as $p_T\sim Q_s(x)$. In Fig.~\ref{fig:comparpp} (left) we present our predictions for the low-$p_{T}$ region compared to the data collected by the CMS and ATLAS experiments at 2.76 and 7~TeV. One can see that all three CD models are able to describe the data in the four rapidity bins. Comparing these results to those presented in Fig.~3 of Ref.~\cite{Benic:2019fbj}, labeled as CGC in Fig.~\ref{fig:comparpp}, one finds a similar good description of the data, however Ref.~\cite{Benic:2019fbj} assumes a $K$-factor while our results are parameter-free in all rapidity bins. Based on this evidence, we also present our predictions for the prompt photon production at 13~TeV in three rapidity bins, which demonstrates the need for more data in order to confirm the good description provided by the CD models at a lower $p_{T}$ range.

\begin{figure*}[t]
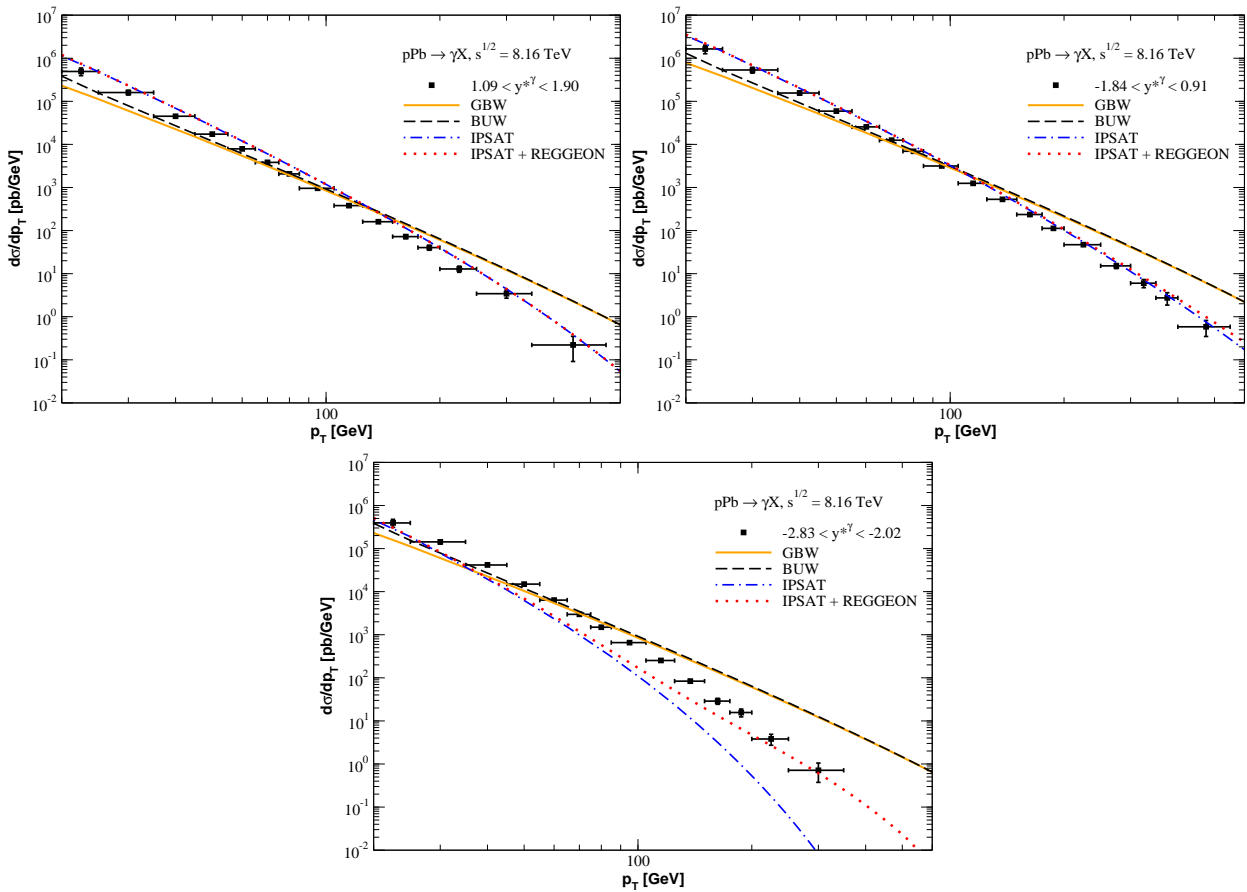

\begin{tabular}{cc}
\includegraphics[scale=0.35]{prompt_pA816_ATLAS1.eps}  \includegraphics[scale=0.35]{prompt_pA816_ATLAS2.eps} \\
\includegraphics[scale=0.35]{prompt_pA816_ATLAS3.eps}  
\end{tabular}
\caption{Differential cross sections for prompt photon production in $pPb$ collision at $\sqrt{s} = 8.16$~TeV and considering three rapidity bins. The predictions are obtained using three phenomenological CD models and compared to the experimental data from ATLAS experiment \cite{aaboud1}. Reggeon contribution is also investigated (see text).}
\label{pA816}
\end{figure*}

In the following we present the results for prompt photon production in $pPb$ collisions at $\sqrt{s} = 8.16$~TeV, where the differential cross section in terms of $p_T$ is shown in Fig.~\ref{pA816}. In this case, the experimental results are obtained taking into account three different rapidity bins: $1.09 < y^{*\,\gamma} < 1.90$, $-1.84 < y^{*\,\gamma} < 0.91$, and $-2.83 < y^{*\,\gamma} < -2.02$. The $p_T$ spectrum for the first bin, $y^{\gamma}\approx 1.5$, is probing $x_2\leq 1.3\times 10^{-2}$ and in the backward rapidity bin ($y^{\gamma} \approx -2.4$) large $x$ is probed, $x_2\sim 0.5$. As a remark, the ATLAS data covers the region between small and large $x$ (where the threshold is taken as $x\simeq 10^{-2}$). The theoretical predictions are compared to the experimental data from the ATLAS detector \cite{aaboud1}. For $p_T < 50$~GeV, the GBW and BUW models give predictions slightly below the experimental data points. In this case the nuclear effects are introduced by GS property as discussed in previous section. However, in the kinematic range $50 < p_T < 100$~GeV such models have a better description of the data. At $p_T > 100$~GeV, the results strongly deviate from the experimental measurements. Once again, the IPSAT model does a good description at large $p_T$ in comparison to the GBW and BUW parametrizations, however IPSAT does not describe data in the negative rapidity bin $-2.83 < y^{*\,\gamma} < -2.02$, for which IPSAT accounts a nuclear correction coming from the GG shadowing. It is surprising that QCD CD models still describe part of the $p_T$ spectrum correctly despite the large $x_2$ values involved in the measured kinematic range. In the case of IPSAT, the large $x$ threshold is given by $(1-x_2)^6$ in the input for the gluon distribution at initial scale. 

The rapidity range covered by the ATLAS experiment in $pA$ case leads to still larger values of $x_2$, mostly in the very backward direction. Hence, we will investigate the role played by the valence quark contribution to the process within the CD framework. In order to do so, we add a Reggeon contribution to the CD amplitude, Eq.~(\ref{ipsat1}), based on Refs.~\cite{Betemps:2003je,Kopeliovich:2001hf}, which results in
\begin{eqnarray}
\sigma_{I\!\!R}^{nuc} (x_2,\vec{r};A) &=& 2\int d^2b\,N_{I\!\!R}^A(x_2,r,b;A),\\\nonumber
N_{I\!\!R}^A(x_2,r,b;A)& = &  N_0r^2x_2^{0.425}(1-x_2)^3\left[\xi_VR_V(x_2,\mu^2) T_A(b)\right],\\
\label{reggeon}
\end{eqnarray}
with $R_V$ is the nuclear ratio for valence quarks (taken from EPPS16 parametrization \cite{Eskola:2016oht}). The parameter $N_0=0.18$ is determined from the $p_T$ spectra for $pp$ collisions at 8~TeV \cite{Aad:2016xcr}. The calculation using the IPSAT model plus a Reggeon contribution is shown in Fig.~\ref{pA816} labeled by dotted lines. The parameter $\xi_V=2/3$ quantifies our ignorance about the actual normalization for the nuclear effects in the Reggeon sector. We confirm that the valence quark contribution plays an important role only in large backward photon pseudorapidities: it modifies spectra at large $p_T$ and will be significant in the prediction for the nuclear modification factor. Notice that the valence quark dependence presented above is model dependent and other phenomenological proposals can be considered.

It is timely to discuss now the uncertainty coming from the model for the nuclear saturation scale used in the GS predictions. Quantitatively, the nuclear saturation scale obtained from Eq.~(\ref{qs2A}) is  $Q_{s,Pb}^2\approx 3 Q_{s,p}^2$ for a Lead nucleus ($A=208$). The value of $Q_{s,A}$ can change whether distinct treatments of the nuclear collision geometry are considered. As an example, using a local saturation scale, $Q_s^2(x,b)=Q_s^2(x,b=0)T_A(b)$, with $T_A$ being the nuclear thickness function and a Gaussian $b$-profile, the relation between $Q_{s,A}$ and $Q_{s,p}$ is found in Ref.~\cite{Salazar:2019ncp}. In the hard sphere approximation for the nuclear density $\rho_A$, one has $Q_{s,A}^2=3A(R_p/R_A)^2Q_{s,p}^2$, which produces $Q_{s,Pb}^2\approx 2.3 Q_{s,p}^2$. Therefore, the typical theoretical uncertainty on the determination of the saturation scale compared to the proton one is $\sim$20$\%$. The ATLAS measurement in $pPb$ collisions in forward rapidities is scanning values of $x_2$ in the range $10^{-3}\lesssim x_2\lesssim 10^{-2}$ on the measured $p_T$ range. This implies in a nuclear saturation scale having values of order $0.8\lesssim Q_{s,Pb}^2\lesssim 1.4$~GeV$^2$, which demonstrates that $p_T\gg Q_{s,A}$ as in the proton case. 

\begin{figure*}[t]
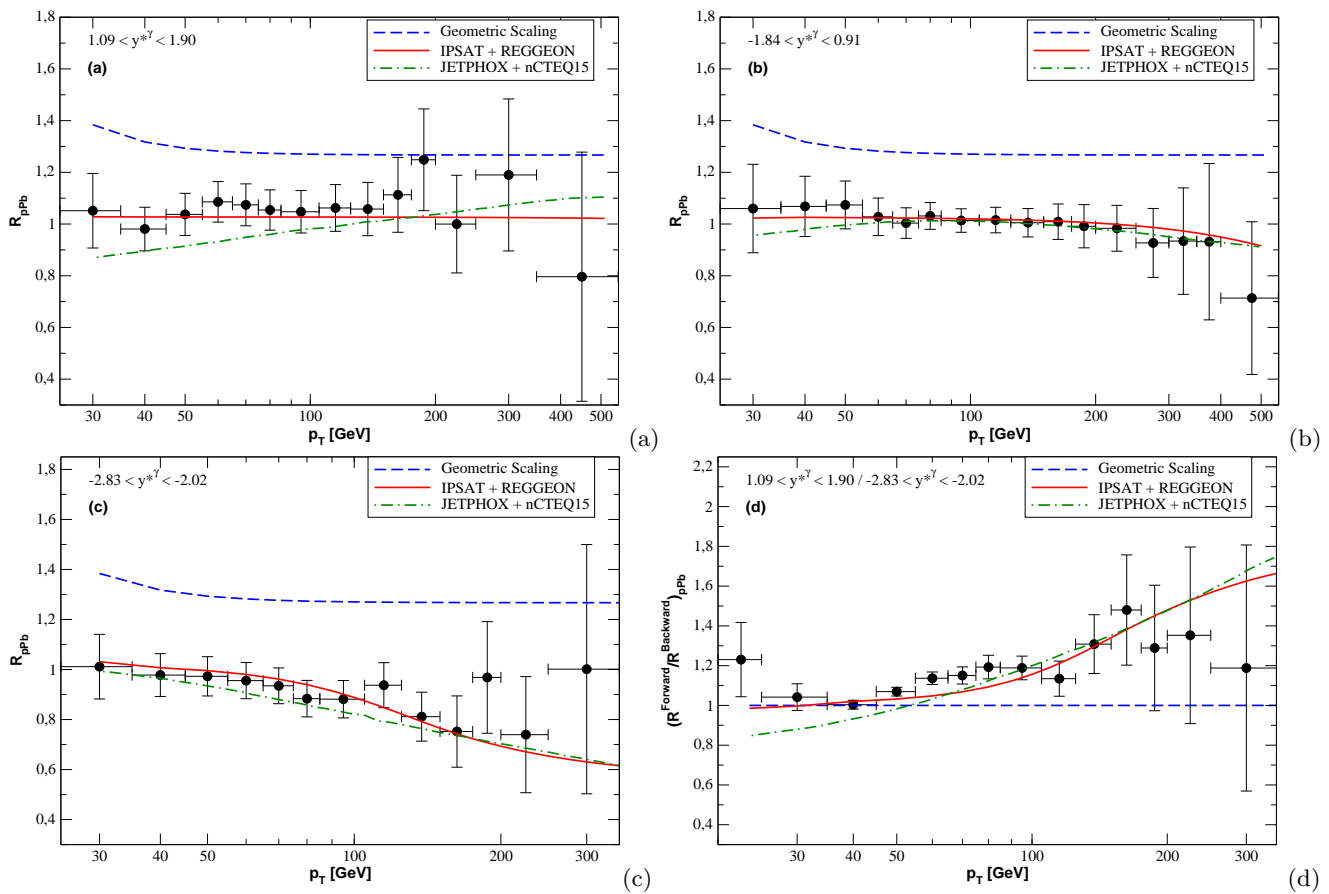

\begin{tabular}{cc}
\includegraphics[scale=0.35]{rpPb_109_y_190.eps} (a) \includegraphics[scale=0.35]{rpPb_184_y_091.eps} (b) \\
\includegraphics[scale=0.35]{rpPb_283_y_202.eps}(c)  \includegraphics[scale=0.35]{rpPb_forw_back.eps}  (d)
\end{tabular}
\caption{Nuclear modification factor $R_{pPb}$ for isolated photons as a function of photon transverse momentum, $p_T$, shown for different center-of-mass rapidity, $y^{\gamma *}$ at $\sqrt{s} = 8.16$~TeV. The predictions are obtained using two phenomenological CD models (BUW and IPSAT) and compared to the NLO pQCD calculation including direct and fragmentation contributions to the cross section (with nuclear PDF nCTEQ15).}
\label{RpA}
\end{figure*}

Finally, a comparison of the predictions for the nuclear modification factor, $R^{\gamma}_{pA}(y,p_T)$, is done in what follows. In Fig.~\ref{RpA} the nuclear ratios are calculated as a function of photon transverse energy in the three $y^{\gamma *}$ regions as in Fig.~\ref{pA816}. We present the two possible ways to include nuclear effects: (i) GS property (dashed lines), Eqs.~(\ref{prescr})-(\ref{qs2A}), where the nuclear dependence is absorbed into the nuclear saturation scale, and (ii) GG shadowing (solid lines), Eq.~(\ref{ipsat2}), where nuclear dependence results from the multiple scattering of CDs off nuclei. Here, the GS approach is applied using the BUW parametrization for dipole-proton amplitude (similar results are obtained by using GBW amplitude). In the GG approach we took the IPSAT model for the dipole-proton amplitude and add the Reggeon contribution, Eq.~(\ref{reggeon}).   The ratio is computed as follows,
\begin{eqnarray}
R^{\gamma}_{pA}(p_T) = \frac{d\sigma(p+Pb\rightarrow \gamma + X)/dp_T}{A\cdot d\sigma (p+p\rightarrow \gamma +X)/dp_T},
\end{eqnarray}
where the photon rapidity is integrated over a given interval. Our calculation corresponds to direct production and the fragmentation contribution is not included. The Reggeon contribution has been added to the CD approach as discussed before (in $pp$ case, we replace $T_A(b)\rightarrow T_p(b)$ by $\xi_VR_V=1$ in Eq.~\ref{reggeon}). For the sake of comparison, we present also the full NLO pQCD calculation (Jetphox Monte Carlo) of the direct and fragmentation contributions to the cross-sections with the nCTEQ15 nuclear PDF set  \cite{aaboud1} (dot-dashed lines).

It is seen that at forward photon rapidities (Fig.~\ref{RpA}-a), the measured nuclear modification factor value is consistent with unity, indicating that nuclear effects are negligible. Notice that in this kinematic region, Reggeon contribution is negligible as demonstrated before. Both GS and GG approaches predict quite small nuclear effect. In GS, this can be easily traced back to the semi-analytical result in Eq.~(\ref{dsdpapprox}). As the rapidity dependence is factorized out in the $f(y)$ function, the nuclear ratio from GS approach is given by (with $x_T=2p_T/\sqrt{s}$),
\begin{eqnarray}\nonumber
R^{\gamma}_{pA}&=& \frac{d\sigma (pA)/dyd^2p_T}{A\cdot d\sigma (pp)/dyd^2p_T} \sim \frac{S_AQ_{s,p}^2(x_T/2)Q_{s,A}^2(x_T/2)}{AS_p[Q_{s,p}^2(x_T/2)]^2}\\
&\approx & \frac{S_A}{AS_p}\left(\frac{AS_p}{S_A}  \right)^{\Delta}=\left( \frac{A\pi R_p^2}{\pi R_A^2} \right)^{\frac{(1-\delta)}{\delta}},
\end{eqnarray}
where $\Delta=1/\delta=1+(1-\delta)/\delta\simeq1+0.27$. Numerically, this would give a value $R_{pA}\simeq 1.3$ (in BUW case) for any value of $p_T$. This behavior is clearly seen in the full numerical calculation, including the order of the magnitude for the nuclear ratio. Concerning the GG approach, in the small $r$ approximation valid here the eikonals shown in Eqs.~(\ref{ipsat1}) and (\ref{ipsat2}) can be both expanded as $N_p\approx (\pi^2\alpha_s/2N_c)r^2 xgT_p(b)$ and $N_A\approx (\pi^2\alpha_s/2N_c)r^2 xgT_A(b)$. By using the normalization for the proton profile function, $\int d^2\vec{b}T_p(b)=1$, and for the nuclear thickness function, $\int d^2\vec{b}T_A(b)=A$, then $\sigma_{dip}^{nuc}=A\sigma_{dip}$ and the predicted ratio (without Reggeons) is $R_{pA}\approx 1$. The situation remains the same in Fig.~\ref{RpA}-b with a tiny contribution from Reggeons at low and large $p_T$. The GS and GG prediction are very close to those from Jetphot Monte Carlo with nCTEQ15. On the other hand, in backward rapidities (Fig.~\ref{RpA}-c) the GG prediction is dominated by Reggeon contribution at large $p_T$. Therefore, the nuclear effect is driven by the valence quark nuclear ratio, $R_V(x_2)$. The average rapidity in this case is $\langle \eta^{\gamma} \rangle = -2.42$ and the average $x_2$ coverage $\langle  x_2 \rangle=[0.04{,}0.42]$, which scans also the EMC-effect region. The observed suppression in the ratio is also the expected behavior in NLO pQCD calculations at backward rapidities \cite{Arleo:2011gc,Arleo:2007js},  where $R_{pA}^{y<0}(x_T)\simeq R_{F_2}^A(x_Te^{-y})$.

Now moving to the production ratio at forward-over-backward rapidity, we compute $R_{\mathrm{FB}}$ which is a quantity that better isolate the size of nuclear effects. If it is computed for symmetric rapidity range,  $R_{\mathrm{FB}}$ is independent of the reference to $pp$ collisions. It is computed in the following way by ATLAS,
\begin{eqnarray}
R_{\mathrm{FB}}(p_T) = \frac{R^{\gamma}_{pA}(1.09<y^{\gamma *}<1.90)}{R^{\gamma}_{pA}(-2.83<y^{\gamma *}<-2.02)},
\end{eqnarray}
and it is presented in Fig.~\ref{RpA}-d. At large $p_T$ the GG calculation including Reggeons is quite similar to the NLO pQCD result, meaning that the valence quark contribution is driven the ratio as a function of transverse momentum.  It was demonstrated in Refs.~\cite{Arleo:2011gc,Arleo:2007js} that the FB ratio in symmetric case $y\rightarrow -y$, behaves like $R_{\mathrm{FB}}=R_{pA}(y)/R_{pA}(-y)=R_g^A(x_Te^{-y})/R_{F_2}^A(x_Te^y)$. This trend is partially followed by data and by the theoretical prediction including valence quark contribution.

As a final comment, it would be timely to compare our calculations as those performed in Ref.~\cite{Ducloue:2017kkq} within the CGC framework using CD cross sections solved from the running coupling Balitsky-Kovchegov (BK) evolution equation and also  predictions in Ref.~\cite{Benic:2016uku} using CGC formalism at level NLO and nuclear UGD. This will be postponed for a future investigation.

\section{Region of validity}
\label{valid}

Discussing on the $x$-values probed by the CMS and ATLAS detectors in the measured $p_T$ ranges, we see that, at central rapidities at 13~TeV, CMS covers $8\times 10^{-3} \lesssim x_2 \leq 7\lesssim 10^{-2}$ and in its extreme forward bin ($y^{\gamma}\approx 2.3$) one reaches $8\times 10^{-5}  \lesssim x_2 \lesssim 6\times 10^{-4}$, while similar ranges of $x$ are covered by the ATLAS detector. These values can be translated to the corresponding saturation scale, Eq.~(\ref{param}), in those rapidity regions. For CMS measurements, one has $0.16\lesssim Q_s^2(x_2) \lesssim 0.27$~GeV$^2$ at midrapidities and $0.5 \lesssim Q_s^2(x_2) \lesssim 0.85$ in the very forward rapidities. Given these typical values of $x$, it means that an important part of $\alpha$ integration is probing $F_2(x_1, Q^2)$ at relatively small $x$ at central rapidities, where $x_1=x_2$. Hence, it is clear that the proton saturation scale in the $p_T$ range measured by the CMS and ATLAS detectors is quite smaller than the transverse momenta, $Q_s^2\ll~p_T^2$, and the color transparency approximation for the CD amplitude is quite well justified. The situation will change only for measurements that reach mild to small ranges in transverse momentum, namely $p_T\lesssim 10$~GeV.

Some words about the validity of CD approach are in order. Although both valence and sea quarks in the projectile are taken into account through the proton $F_2$ in Eq.~(\ref{xsec}), the CD picture accounts only for Pomeron exchange from the target. Therefore, in principle it is well suited for  small $x_2$. The CD expression, Eq.~(\ref{xsec}), is valid for any value of $x_1$ as it enters in the proton structure function $F_2(x_t,Q^2)$. In the $\alpha$-integration in Eq.
~(\ref{hank}), one has $x_1<x_t<1$ (with $x_t=x_1/\alpha$) and we are using a parametrization for the structure function \cite{adeva} valid in the range $8\times 10^{-4}<x_t<0.7$. In Ref.~\cite{mm} the updated ALLM parametrization was used (it covers $3\times 10^{-6} < x_t < 0.85$) and the numerical results are practically the same. We have checked that the output is the same in a large range of $p_T$ by using Ref.~\cite{adeva} or ALLM2007 parametrization. Concerning the $x_2$ range in our numerical calculations, the GBW and BUW dipole cross sections have been multiplied by a factor $(1-x_2)^n$ (with $n=7$) in order to take into account the large-$x$ behavior of cross sections. 

For the IPSAT model, the threshold factor is already included in the parametrization for the gluon PDF at the initial scale. The role played by this threshold factor for prompt photon-spectra within the CD approach has been investigated in Ref.~\cite{mm}. In Ref.~\cite{Betemps:2003je} one of us investigated the extrapolation of the dipole approach to very large-$x$ by introducing a Reggeon contribution (in the context of Drell-Yan production). This Reggeon part is proportional to the valence quark content of the target, meaning that it is negligible at the RHIC and the LHC energies, although it is important in order to obtain a good description of the low-energy CERN ISR data. The Reggeon contribution can be perceptible at very backward rapidities even at the LHC energies and we will come back to this point in $pA$ case.

For sake of illustration, the extrema $x_{2}=(p_{T}/\sqrt{s})e^{-y}$ range for CMS and ATLAS $pp$ data at 13~TeV and ATLAS $p$A 8.16~TeV are presented at Tab.~\ref{tabex}. Ones sees that the ATLAS $p$A data in the rapidity range of $-2.83<y<-2.02$ probes rather larger $x$ values, which is evident in the comparison between theory-experiment in Fig.~\ref{pA816}. To cover additional large-$x$ contributions, we have introduced a Reggeon contribution in $p$A case in order to improve the data description. As expected, the Reggeon contribution is extremely small unless at very large backward rapidities, showing that the CD approach gives reasonable results for large $x$, which is not restrict to $x$ values below 10$^{-2}$ like other approaches such as CGC.

\begin{table}
\centering
\begin{tabular}{cc|c}
\hline\hline
$x_{2,min}$ & $x_{2,max}$ & Run \\
\hline
0.0067 & 0.0673 & CMS 13~TeV \\
0.0035 & 0.0302 & CMS 13~TeV \\
0.0018 & 0.0140 & CMS 13~TeV \\
0.0012 & 0.0058 & CMS 13~TeV \\
\hline
0.0058 & 0.1    & ATLAS 13~TeV \\
0.0027 & 0.0548 & ATLAS 13~TeV \\
0.0017 & 0.0161 & ATLAS 13~TeV \\
0.0010 & 0.0103 & ATLAS 13~TeV \\
\hline
0.0004	& 0.0176 & ATLAS 8.16~TeV \\
0.0010	& 0.3477 & ATLAS 8.16~TeV \\
0.0197	& 0.5911 & ATLAS 8.16~TeV \\
\hline\hline
\end{tabular}
\label{tabex}
\caption{Extreme values of $x_{2}$ probed within the kinematic regions of the CMS and ATLAS detectors in both $pp$ and $p$A collisions at the LHC.}
\end{table}

\section{Summary} 
\label{conc}
We investigate the prompt photon production at small $x$ in $pp$ and $pA$ collisions at the LHC energies at different rapidity bins. We show that direct photon production can be formulated in the QCD CD framework without any free parameter. In particular, we employ three dipole cross section models determined by recent phenomenological analysis of DIS data available from DESY-HERA. The predictions for $pp$ and $pA$ reactions have demonstrated that in the low-$p_{T}$ range we can not completely distinguish between GBW, BUW, and IPSAT models. Nonetheless, the IPSAT results provide a better description of the data at the high-$p_T$ range compared to the other CD models based on fixed or running effective anomalous dimension. The particular case of $pPb$ collisions at the LHC have show that the result with the IPSAT model have a good agreement with the ATLAS data up to mid-rapidities, where the the values of $x$ probed in this range are small. At more forward rapidities, the results with IPSAT are beyond its limit of validity (larger $x$), showing that the predictions undershoot the data as expected.

Therefore, our results encourage for additional improvements that may be taken into account to refine the corresponding phenomenology at the large $p_T$ spectrum if new data from the LHC energy regime become available. Furthermore, we propose that future measurements of prompt photon production in $pp/pA/AA$ collisions may be performed at the current/future colliders, since these data could be a valuable tool to analyze the CD models as wells as constrain high energy QCD dynamics effects such as saturation physics in kinematic domain not yet explored.

\section*{Acknowledgments}

This work was partially financed by the Brazilian funding agencies CAPES, CNPq, and FAPERGS. This study was financed in part by the Coordena\c{c}\~ao de Aperfei\c{c}oamento de Pessoal de N\'{\i}vel Superior - Brasil (CAPES) - Finance Code 001. GGS
acknowledges funding from the Brazilian agency Conselho Nacional
de Desenvolvimento Científico e Tecnológico (CNPq) with grant
CNPq/313342/2017-2.


\begin{thebibliography}{99}

\bibitem{atlas} G.~Aad {\it et al.} [ATLAS Collaboration], Phys.\ Rev.\ D {\bf 83}, 052005 (2011), arxiv:1012.4389 [hep-ex]; Phys.\ Lett.\ B {\bf 706}, 150 (2011), arxiv:1108.0253 [hep-ex]; Phys.\ Rev.\ D {\bf 89}, 052004 (2014), arxiv:1311.1440 [hep-ex].

\bibitem{cms} V.~Khachatryan {\it et al.} [CMS Collaboration], Phys.\ Rev.\ Lett.\  {\bf 106}, 082001 (2011), arxiv:1012.0799 [hep-ex]; Phys.\ Rev.\ D {\bf 84}, 052011 (2011), arxiv:1108.2044 [hep-ex]. 

\bibitem{cdf} T.~Aaltonen {\it et al.} [CDF Collaboration], Phys.\ Rev.\ D {\bf 80}, 111106 (2009), arxiv:0910.3623 [hep-ex]. 

\bibitem{d0} V.~M.~Abazov {\it et al.} [D0 Collaboration], Phys.\ Lett.\ B {\bf 639}, 151 (2006), arxiv:hep-ex/0511054 [hep-ex]; Phys.\ Lett.\ B {\bf 725}, 6 (2013), arxiv:1301.4536 [hep-ex].

\bibitem{bayatian} G.~L.~Bayatian {\it et al.} [CMS Collaboration], J.\ Phys.\ G {\bf 34}, 995 (2007).

\bibitem{aad} G.~Aad {\it et al.} [ATLAS Collaboration], Phys.\ Rev.\ D {\bf 90}, 112015 (2014), arxiv:1408.7084 [hep-ex].

\bibitem{pasechnik} R.~Pasechnik and M.~Sumbera, Universe {\bf 3}, 7 (2017), arxiv:1611.01533 [hep-ph].

\bibitem{acharya} S.~Acharya {\it et al.} [ALICE Collaboration], Phys.\ Rev.\ C {\bf 99}, 024912 (2019), arxiv:1803.09857 [nucl-ex].

\bibitem{gordon} L.~E.~Gordon and W.~Vogelsang, Phys.\ Rev.\ D {\bf 50}, 1901 (1994).

\bibitem{frixione} S.~Frixione, Phys.\ Lett.\ B {\bf 429}, 369 (1998), arxiv:hep-ph/9801442 [hep-ph].

\bibitem{helenius} I.~Helenius, K.~J.~Eskola and H.~Paukkunen, JHEP {\bf 05}, 030 (2013), arxiv:1302.5580 [hep-ph].

\bibitem{steffen} F.~D.~Steffen and M.~H.~Thoma, Phys.\ Lett.\ B {\bf 660}, 604 (2008), arxiv:hep-ph/0103044 [hep-ph].

\bibitem{rapp} R.~Rapp, H.~van Hees and M.~He, Nucl.\ Phys.\ A {\bf 931}, 696 (2014), arxiv:1408.0612 [nucl-th].

\bibitem{kop} B.~Z.~Kopeliovich, A.~H.~Rezaeian, H.~J.~Priner and I.~Schmidt, Phys. Lett. B {\bf 653}, 210 (2007), arxiv:0704.0642 [hep-ph].

\bibitem{kop1} B.~Z.~Kopeliovich, A.~V.~Tarasov and A.~Schafer, Phys.\ Rev.\ C {\bf 59}, 1609 (1999), arxiv:hep-ph/9808378 [hep-ph].

\bibitem{kop2} B.~Z.~Kopeliovich, proc.\ of the workshop Hirschegg '95:
Dynamical Properties of Hadrons in Nuclear Matter, Hirschegg January
16-21, 1995, ed. by H. Feldmeyer and W. N\"orenberg, Darmstadt, 1995, p. 102, arxiv:hep-ph/9609385.

\bibitem{Kopeliovich:2007fv} 
B.~Z.~Kopeliovich, H.~J.~Pirner, A.~H.~Rezaeian and I.~Schmidt,
Phys.\ Rev.\ D {\bf 77}, 034011 (2008), arxiv:0711.3010 [hep-ph].

\bibitem{Benic:2018hvb} 
S.~Benic, K.~Fukushima, O.~Garcia-Montero and R.~Venugopalan,
Phys.\ Lett.\ B {\bf 791}, 11 (2019), arxiv:1807.03806 [hep-ph].

\bibitem{Benic:2019fbj} 
S.~Benic, K.~Fukushima, O.~Garcia-Montero and R.~Venugopalan,
MDPI Proc.\  {\bf 10}, no. 1, 33 (2019).

\bibitem{Benic:2016uku} 
S.~Benic, K.~Fukushima, O.~Garcia-Montero and R.~Venugopalan,
JHEP {\bf 01}, 115 (2017), arxiv:1609.09424 [hep-ph].


\bibitem{Ducloue:2017kkq} 
B.~Ducloue, T.~Lappi and H.~Mantysaari,
Phys.\ Rev.\ D {\bf 97}, no. 5, 054023 (2018), arxiv:1710.02206 [hep-ph].

\bibitem{Kopeliovich:2009yw} 
B.~Z.~Kopeliovich, E.~Levin, A.~H.~Rezaeian and I.~Schmidt,
Phys.\ Lett.\ B {\bf 675}, 190 (2009), arxiv:0902.4287 [hep-ph].

\bibitem{Rezaeian:2009it} 
A.~H.~Rezaeian and A.~Schafer,
Phys.\ Rev.\ D {\bf 81}, 114032 (2010), arxiv:0908.3695 [hep-ph].

\bibitem{Krelina:2016hkr} 
M.~Krelina, E.~Basso, V.~P.~Goncalves, J.~Nemchik and R.~Pasechnik,
EPJ Web Conf.\  {\bf 120}, 03006 (2016).

\bibitem{JalilianMarian:2012bd} 
J.~Jalilian-Marian and A.~H.~Rezaeian,
Phys.\ Rev.\ D {\bf 86}, 034016 (2012), arxiv:1204.1319 [hep-ph].

\bibitem{Kovner:2014qea} 
A.~Kovner and A.~H.~Rezaeian,
Phys.\ Rev.\ D {\bf 90}, no. 1, 014031 (2014), arxiv:1404.5632 [hep-ph].

\bibitem{Rezaeian:2016szi} 
A.~H.~Rezaeian,
Phys.\ Rev.\ D {\bf 93}, no. 9, 094030 (2016), arxiv:1603.07354 [hep-ph].

\bibitem{Kovner:2015rna} 
A.~Kovner and A.~H.~Rezaeian,
Phys.\ Rev.\ D {\bf 92}, no. 7, 074045 (2015), arxiv:1508.02412 [hep-ph].

\bibitem{Benic:2017znu} 
S.~Beni\'c and A.~Dumitru,
Phys.\ Rev.\ D {\bf 97}, no. 1, 014012 (2018), arxiv:1710.01991 [hep-ph].

\bibitem{vic} V.~P.~Goncalves, Y.~Lima, R.~Pasechnik and M.~Sumbera, arXiv:2003.02555 [hep-ph].

\bibitem{goharipour} M.~Goharipour and S.~Rostami, Phys.\ Rev.\ C {\bf 99}, 055206 (2019), arxiv:1808.05639 [hep-ph].

\bibitem{campbell} J.~M.~Campbell, J.~Rojo, E.~Slade and C.~Williams, Eur.\ Phys.\ J.\ C {\bf 78}, 470 (2018), arxiv:1802.03021 [hep-ph].

\bibitem{roy} K.~Roy and R.~Venugopalan, JHEP {\bf 05}, 013 (2018), arxiv:1802.09550 [hep-ph].  

\bibitem{mm} M.~V.~T.~Machado and C.~B.~Mariotto, Eur.\ Phys.\ J.\ C {\bf 61}, 871 (2009), arxiv:0809.1884 [hep-ph].

\bibitem{favartmm} L. Favart and M.~V.~T.~Machado, Eur.\ Phys.\ J.\ C {\bf 29}, 365 (2003), arxiv:hep-ph/0302079 [hep-ph].

\bibitem{Machado:2005ez} 
M.~V.~T.~Machado,
Eur.\ Phys.\ J.\ C {\bf 47}, 365 (2006), arxiv:hep-ph/0512264 [hep-ph].

\bibitem{Goncalves:2006yt} 
V.~P.~Goncalves, M.~S.~Kugeratski, M.~V.~T.~Machado and F.~S.~Navarra,
Phys.\ Lett.\ B {\bf 643}, 273 (2006), arxiv:hep-ph/0608063 [hep-ph].

\bibitem{armesto} N.~Armesto, C.~A.~Salgado and U.~A.~Wiedemann, Phys.\ Rev.\ Lett.\  {\bf 94}, 022002 (2005), arxiv:hep-ph/0407018 [hep-ph].

\bibitem{kop3} B.~Z.~Kopeliovich, J.~Raufeisen and A.~V.~Tarasov, Phys.\ Lett.\ B {\bf 503}, 91 (2001), arxiv:hep-ph/0012035 [hep-ph].

\bibitem{adeva} B.~Adeva {\it et al.}, Phys.\ Rev.\ D {\bf 58}, 112001 (1998).

\bibitem{Kopeliovich:2001hf} 
B.~Z.~Kopeliovich, J.~Raufeisen, A.~V.~Tarasov and M.~B.~Johnson,
Phys.\ Rev.\ C {\bf 67}, 014903 (2003), arxiv:hep-ph/0110221 [hep-ph].

\bibitem{field} R.~D.~Field, Applications of Perturbative QCD, Perseus Books, Reading, Massachusetts, 1989.

\bibitem{Betemps:2003je} 
M.~A.~Betemps, M.~B.~G.~Ducati, M.~V.~T.~Machado and J.~Raufeisen,
Phys.\ Rev.\ D {\bf 67}, 114008 (2003), arxiv:hep-ph/0303100 [hep-ph].

\bibitem{Raufeisen:2002zp} 
J.~Raufeisen, J.~C.~Peng and G.~C.~Nayak,
Phys.\ Rev.\ D {\bf 66}, 034024 (2002), arxiv:hep-ph/0204095 [hep-ph].


\bibitem{kop4} A.~B.~Zamolodchikov, B.~Z.~Kopeliovich and L.~I.~Lapidus, JETP Lett. {\bf 33}, 595 (1981).

\bibitem{gbw} K.~J.~Golec-Biernat and M.~Wusthoff, Phys.\ Rev.\  D {\bf 59}, 014017 (1998), arxiv:hep-ph/9807513 [hep-ph].

\bibitem{gbwfit} K.~Golec-Biernat and S.~Sapeta, JHEP {\bf 03}, 102 (2018), arxiv:1711.11360 [hep-ph].

\bibitem{buw} D.~Boer, A.~Utermann and E.~Wessels, Phys.\ Rev.\ D {\bf 77}, 054014 (2008), arxiv:0711.4312 [hep-ph].

\bibitem{ipsat} H.~Kowalski and D.~Teaney, Phys.\ Rev.\ D {\bf 68}, 114005 (2003), arxiv:hep-ph/0304189 [hep-ph].

\bibitem{ipsatfit} A.~H.~Rezaeian, M.~Siddikov, M.~Van de Klundert and R.~Venugopalan, Phys.\ Rev.\ D {\bf 87}, 034002 (2013), arxiv:1212.2974 [hep-ph].

\bibitem{Ben:2017xny} 
F.~G.~Ben, M.~V.~T.~Machado and W.~K.~Sauter,
Phys.\ Rev.\ D {\bf 96}, 054015 (2017), arxiv:1701.01141 [hep-ph].

\bibitem{dEnterria:2012kvo}
D.~d'Enterria and J.~Rojo,
Nucl. Phys. B \textbf{860}, 311-338 (2012), arxiv:1202.1762 [hep-ph].

\bibitem{Catani:2002ny}
S.~Catani, M.~Fontannaz, J.~Guillet and E.~Pilon,
JHEP \textbf{05}, 028 (2002), arxiv:hep-ph/0204023 [hep-ph].

\bibitem{sirunyan} A.~M.~Sirunyan {\it et al.} [CMS Collaboration], Eur.\ Phys.\ J.\ C {\bf 79}, 20 (2019), arxiv:1807.00782 [hep-ex].

\bibitem{aaboud} M.~Aaboud {\it et al.} [ATLAS Collaboration], Phys.\ Lett.\ B {\bf 770}, 473 (2017), arxiv:1701.06882 [hep-ex], .

\bibitem{aaboud1} M.~Aaboud {\it et al.} [ATLAS Collaboration], Phys.\ Lett.\ B {\bf 796}, 230 (2019), arxiv:1903.02209 [nucl-ex].

\bibitem{Eskola:2016oht}
K.~J.~Eskola, P.~Paakkinen, H.~Paukkunen and C.~A.~Salgado,
Eur. Phys. J. C \textbf{77}, no.3, 163 (2017), arxiv:1612.05741 [hep-ph].

\bibitem{Aad:2016xcr}
G.~Aad \textit{et al.} [ATLAS],
JHEP \textbf{08}, 005 (2016), arxiv:1605.03495 [hep-ex].

\bibitem{Arleo:2011gc}
F.~Arleo, K.~J.~Eskola, H.~Paukkunen and C.~A.~Salgado,
JHEP \textbf{04}, 055 (2011), arxiv:1103.1471 [hep-ph].

\bibitem{Arleo:2007js}
F.~Arleo and T.~Gousset,
Phys. Lett. B \textbf{660}, 181-187 (2008), arxiv:0707.2944 [hep-ph].

\bibitem{Salazar:2019ncp} 
F.~Salazar and B.~Schenke,
Phys.\ Rev.\ D {\bf 100}, no. 3, 034007 (2019), arxiv:1905.03763 [hep-ph].

\bibitem{Klein-Bosing:2014uaa} 
C.~Klein-Bösing and L.~McLerran,
Phys.\ Lett.\ B {\bf 734}, 282 (2014), arxiv:1403.1174 [nucl-th].

\bibitem{Praszalowicz:2018vfw} 
M.~Praszałowicz,
EPJ Web Conf.\  {\bf 206}, 02002 (2019), arxiv:1812.04524 [hep-ph].

\bibitem{Khachatryan:2019uqn} 
V.~Khachatryan and M.~Praszalowicz,
arXiv:1907.03815 [nucl-th].
  
\end{thebibliography}
\end{document}